\documentclass{styles/svproc}
\usepackage{url}

\usepackage{graphicx}
\graphicspath{{figures}}
\usepackage{makeidx}
\usepackage{hyperref}
\usepackage{blindtext}
\usepackage{xcolor}
\usepackage{enumitem}
\usepackage{amssymb}
\usepackage{pdfpages}
\usepackage{orcidlink}

\newcommand{\beasbox}[1]{\vspace{0.3em}\setlength{\fboxsep}{0.015\linewidth}\noindent\fbox{\parbox{0.96\linewidth}{#1}} \vspace{0.3em}}

\begin{document}
\mainmatter              

\title{Exploring Human-AI Collaboration in Agile: Customised LLM Meeting Assistants}
\titlerunning{Exploring Human-AI Collaboration in Agile}

\author{Beatriz Cabrero-Daniel \orcidlink{0000-0001-5275-8372}$^1$ \and
Tomas Herda  \orcidlink{0009-0005-2912-380X}$^2$ \and \\
Victoria Pichler \orcidlink{0009-0006-4406-127X}$^2$ \and
Martin Eder \orcidlink{0009-0009-7175-4073}$^2$}
\authorrunning{Cabrero-Daniel et al.}
\tocauthor{Beatriz Cabrero-Daniel,
Tomas Herda,
Victoria Pichler,
Martin Eder}
\institute{$^1$University of Gothenburg, \\ $^2$Austrian Post}

\maketitle 

\begin{abstract}
This action research study focuses on the integration of ``AI assistants'' in two Agile software development meetings: the Daily Scrum and a feature refinement, a planning meeting that is part of an in-house Scaled Agile framework. We discuss the critical drivers of success, and establish a link between the use of AI and team collaboration dynamics. We conclude with a list of lessons learnt during the interventions in an industrial context, and provide a assessment checklist for companies and teams to reflect on their readiness level. This paper is thus a road-map to facilitate the integration of AI tools in Agile setups.

\end{abstract}

\section{Introduction}

Team collaboration and meetings are an essential part of any software development organisation, but they are challenging to organise and manage. Often, meetings do not follow guidelines or run into issues that affect their efficiency and productivity, or delay decision-making~\cite{hamdy20}. Sometimes, the guidelines themselves burden the development teams and need to be adapted. The Post-Rolling Refinement Model (PRIME), an in-house designed Scaled Agile framework proposed by the \textit{Austrian Post}~\cite{prime}, aims to reduce this burden.

Microsoft sparked in 2019 the debate on how Artificial Intelligence (AI) could further improve meetings by automating tasks and retrieving information before, during, and after them~\cite{microsoft2023}. That could, among other benefits, improve the flow, save time, increase productivity, or reduce frustration during said meetings~\cite{microsoft2023}. There is a growing body of white and grey literature that recognises the role that AI could have in reducing the organisational burden on the participants, ensuring that meetings are conducted in a more organised and structured manner, or providing insights to improve future meetings. However, there has been little academic attention paid to the role of AI interventions in software development meetings. And, to the best of the authors' knowledge, no research has been conducted to investigate the support of an AI meeting assistant and its interactions with practitioners in a systematic way.

We are adopting action research to explore the use of AI in meetings at \textit{Austrian Post Group IT}, an international postal, logistics and service provider described in detail in Section~\ref{sec:method}. We focus on two of their regular meetings: standard Daily Scrums and feature refinement and planning meetings, part of the PRIME framework~\cite{prime}. The study explores how the use of AI affects the practitioners' meeting experience. By doing so we aim to answer three questions:

\begin{enumerate}[leftmargin=.975cm]
    \item [\textbf{RQ1}] How can AI assist in identifying potential problems and risks in Agile meetings, and provide actionable and useful recommendations?
    \item [\textbf{RQ2}] To what extent do the AI meeting-assistants generate sensible recommendations in the context of real meetings in real time? 
    \item [\textbf{RQ3}] How do users perceive the AI meeting-assistants in terms of user experience, and what impact does it have on overall performance? 
\end{enumerate}

Section~\ref{sec:bg} provides a brief overview of the relevant academic and grey literature on AI for software development tasks. Then, the paper moves on to detail the methodology used in this paper in Section~\ref{sec:method}. Section~\ref{sec:findings} analyses the interventions and evaluation surveys undertaken during this study, and reasons about the design decisions that lead to the final AI meeting-assistants. Finally, Section~\ref{sec:discussion} points out the lessons learnt and how they could be transferred to similar contexts, and Section~\ref{sec:conclusion} highlights important implications for future practice.


\section{Related work} \label{sec:bg}


\subsection{Adapting Agile practices to companies}

Many organisations have already shifted from traditional working processes to Agile. Nevertheless, not all agile teams follow guidelines thoroughly~\cite{howMuchAgile,hamdy20}. This is because they might feel that following all Scrum rules is too time-consuming or that some of the Scrum rules are irrelevant or even outweighing the benefits~\cite{prime}. 

Mortada et al.~discuss several wrong practices in Scrum teams. For example, daily Scrum meeting is supposed to last only 15 minutes, but it often runs longer than that~\cite{hamdy20}. Other examples include not estimating user stories, not having the correct structure for writing the user stories in the backlog, not having a product backlog, not defining the sprint goal at the beginning of the sprint and not ending the sprint with demonstrating the desired deliverable~\cite{hamdy20}. 

Similar problems might also be present in Scaled Agile setups. Previous research has highlighted the role of confusion about roles and responsibilities in creating of unnecessary overhead~\cite{prime}. For example, senior engineers and architects, might want to get more time to understand technical details, which requires separate smaller feature estimation meetings~\cite{prime}. At the same time, developers often do not receive feedback from the right stakeholders at the end of the sprint~\cite{hamdy20}.

For these reasons, some companies have proposed modifications and ad-hoc adaptations to the guidelines. For instance, Austrian Post proposes the Post-Rolling Refinement Model (PRIME)~\cite{prime}. PRIME aims to remove a lot of the bureaucratic overhead, pushes decisions to the Scrum teams, and reduces the number of meetings and the number of mandatory participants~\cite{prime}. 

\subsection{Generative AI in Software Engineering}

Generative AI (GenAI) is a type of AI that can generate different types of content, such as images, text, audio, and 3D models, based on the input it receives. Large Language Models (LLMs), such as \textit{GPT-4}, are currently being used in a wide range of fields, including medicine, economics, software development, academia, and business. 
%
GenAI can also support software engineering~\cite{nguyenduc2023generative}. Managers could use GenAI tools to get recommendations for decision-making, or to automate some interactions with the customers, e.g., using virtual assistants integrated with customer service tools etc~\cite{sarker2022ai}. Organisation's data could also be used as input to the AI tools for the purposes of creating tables, analysing statistics, generating models, and monitoring workflows~\cite{microsoft2023}. 

Microsoft breaks the interventions of AI in before, during, and after team collaboration sessions and meetings and provides insights into how organisations can use AI to retrieve or generate relevant information and resources~\cite{microsoft2023}.
However, there are challenges associated with integrating AI and Agile methodologies, such as the need for specialised technical expertise, and integrating AI into Agile software development processes requires careful consideration of the context~\cite{karac2018we}. Moreover, the most critical challenges are related to human factors, e.g., ensuring that developers have the skills to work with AI and align expectations effectively~\cite{lariosvargas2022,AIact}. As a result, the trade-off between creativity, human oversight, and cyber-security is a critical factor to consider.





\subsection{Prompt Engineering}

Prompt engineering is defined as a set of techniques to improve the inputs or instructions that a user provides for an AI model to get desired outputs~\cite{nguyen2023generative}. Depending on the context that the AI model is used for, designing appropriate prompts is very important to get more accurate results, therefore it is suggested to narrow down the prompts and avoid using too general queries~\cite{hornemalm2023chatgpt}. Mastropaolo et al.~also studied the influence of varying natural language descriptions for Copilot prompts and proved that paraphrasing leads to different quality levels for the generations~\cite{mastropaolo2023robustness}. There are different techniques for prompt engineering including role-prompting, user of triple quotes to separate, trying several times with generating responses, etc~\cite{chen2023unleashing}. 


It is known that it is difficult to balance relevant information retrieval (in this case, generating recommendations) and not overloading the participants~\cite{asthana2023}. These two things can be balanced in the prompt.

\section{Research design} \label{sec:method}

The study aims to understand how practitioners use and perceive the use of AI in meetings, a relatively new phenomenon~\cite{microsoft2023}. We adopted action research to observe both technical and social aspects of AI usage through interventions in \textit{Austrian Post Group IT}'s online meetings. 
The emphasis of this study is not the artefacts produced (provided as Supplementary Materials), but rather the motivations behind design choices, detailed in Section~\ref{sec:findings}. Before that, Section~\ref{sec:methodcontext} provides some context, Section~\ref{sec:methodaction} describes the expectations for both assistants, and Section~\ref{sec:methoddata} explains how surveys and observations were used to draw conclusions and improve the artefact. 


\subsection{Context} \label{sec:methodcontext}


\textit{Austrian Post} is an international postal, logistics and service provider operating in the markets of Austria, where it plays a critical role in the country's infrastructure, eight other countries in Central and Eastern Europe, and Turkey. The development teams, which are part of \textit{Group IT} use broadly accepted frameworks, such as Scrum as well as a in-house designed Scaled Agile framework named PRIME~\cite{prime}. 
More than 520 employees work at \textit{Austrian Post Group IT}, out of which approximately 300 have a similar role as the participants, that belong to 3 out of the 9 development teams at the \textit{Digital Logistics Platform}.

An action team, consisting of both practitioners and researchers, was named to be responsible for planning, executing, and evaluating the research. The selected practitioners were involved in planning and executing actions, besides observing and providing contextualised feedback after each of the interventions. Moreover, they provided the evaluation of the final outcome with their deep knowledge of the context, the Scrum and PRIME frameworks, and the practitioners' way of working.

Complementary, a reference group of practitioners, responsible to give advice and feedback to the action team, and a management team, who is planned to govern the institutionalisation of the proposed changes, were also key to conduct the present study. The goals of these two groups is to evaluate the benefits of AI meeting assistants, as reported in this study, and exploring potential directions for further work, aligned with \textit{Austrian Post}'s strategic goals, e.g., automating repetitive preparation tasks, supporting less experienced developers, creating useful summaries for those that could not attend a meeting, etc.

\subsection{Action} \label{sec:methodaction}

We propose two AI assistants to use before, during and after two Agile software development meetings, as shown in Table~\ref{tab:beforeduringafter}. 
The assistants are instanced by prompting \textit{Azure OpenAI Studio}'s \textit{GPT-4} LLM. The prompts were designed iteratively: first listing the current challenges of each meeting, then refining the prompts and testing them without sharing the generations (silent demos), and finally using the AI assistants with participants, under observation.

\begin{table}[t]
    \centering
    \begin{tabular}{|l|l|p{3cm}|p{3cm}|} \hline 
        \textbf{Intervention} & \multicolumn{1}{p{2.95cm}|}{\textbf{Before meeting}} & \textbf{During meeting} & \textbf{After meeting} \\ \hline \hline 
        PRIME meeting & \multicolumn{2}{p{5.95cm}|}{Creation of slides with identified risks based on current PRIME features board} & - \\ \hline
        Daily Scrum & \multicolumn{1}{p{2.5cm}|}{-} & Real time guidance, ticket management & Post-meeting recommendations \\ \hline
    \end{tabular}
    \caption{AI meeting assistant actions} \label{tab:beforeduringafter}
    \vspace{-0.7cm}
\end{table}


\subsubsection{The Agile Release Train Coach assistant} \label{sec:methodprime}

An AI assistant was designed, with the help of reference team, to help the \textit{Agile Release Train Coach} (a servant leader to the train and support teams in delivering value) prepare and conduct PRIME meetings by helping refine and plan the next PRIME iteration~\cite{prime}.

The \textit{Agile Release Train Coach assistant} is instantiated using a prompt and three spreadsheet files. The files contain information about (i) the features and related children User Stories in the PRIME feature board, (ii) the average velocity of development teams per sprint, and (iii) of the Agile Release Train~\cite{prime}. These files need to be provided to the assistant given that no real-time connection to \textit{Azure DevOps} is yet available for the LLM. The files are automatically embedded by the \textit{Azure AI search} platform in order to be accessed by the LLM~\cite{azuresearch}. Using the embedded files, the assistant provides valuable insights to:
\begin{enumerate}
    \item Limit the risk of teams over-committing (using team's capacity and velocity).
    \item Identify features with no effort value.
    \item Identify features placed in incorrect backlog (based on iteration path).
    \item Check unplanned integration testing efforts.
    \item Identify features where there is a children user story for a team that is not tagged in the feature.
    \item Help plan large features (based on effort points).
    \item Highlight non-estimated and incorrectly estimated features. 

    \item Limit the risk of the Agile release Train over-committing (based on capacity). 
\end{enumerate}

The information contained in the files was manually gathered from \textit{Azure DevOps} and anonymised by the reference team in two-hour-long sessions before each of the PRIME meetings described in Section~\ref{tab:changesprime}. In these sessions, the prompt to instantiate the \textit{Agile Release Train Coach assistant} was improved and the validity of its insights, checked. It is important to note that not only was the AI assistant faster than the reference team at analysing the data, but also made less mistakes than humans. Moreover, as the prompt design improved, as discussed in Sections~\ref{sec:findings}, the time to generate and check the insights went down to 30 minutes. 

\subsubsection{Scrum Team Assistant Tool for the Daily Scrum} \label{sec:methodaily}

Based on the insights of the action and reference team, a second assistant was designed to get real-time recommendations on the meeting progress, and insights on the adherence to the official Scrum guidelines right after the Daily Scrum.

The \textit{Scrum Team Assistant Tool} assistant is instantiated using a prompt and the latest version of the official Scrum Guide, since there is a risk of the LLM retrieving a deprecated version. The prompt gives the general context of the intervention and instructs the LLM on how to act depending on the user inputs. To generate manageable, to-the-point recommendations and align with the expectations of practitioners, the assistant is asked to generate up to 10 words that are ``friendly.'' The prompt concludes with general instructions for the LLM to run the assistant. 
Once instantiated using the prompt in Supplementary Materials, the AI assistant generates different recommendations when:

\begin{itemize}
    \item An individual talks about topics not related to the team's work.
    \item The work that the individual mentions is not visualised in the sprint backlog.
    \item An individual engaged in a detailed discussion about a specific topic.
    \item The impediment that an individual raised is not visualised.
    \item An individual was interrupted by an external circumstance.
    \item An individual does not have any task in ``updated state'' in the backlog.
    \item An individual needs to create a ticket to a specific team.
    \item Any other problem that may occur during the Daily Scrum. 
\end{itemize}

\begin{figure}[t]
    \centering
    \includegraphics[width=\linewidth]{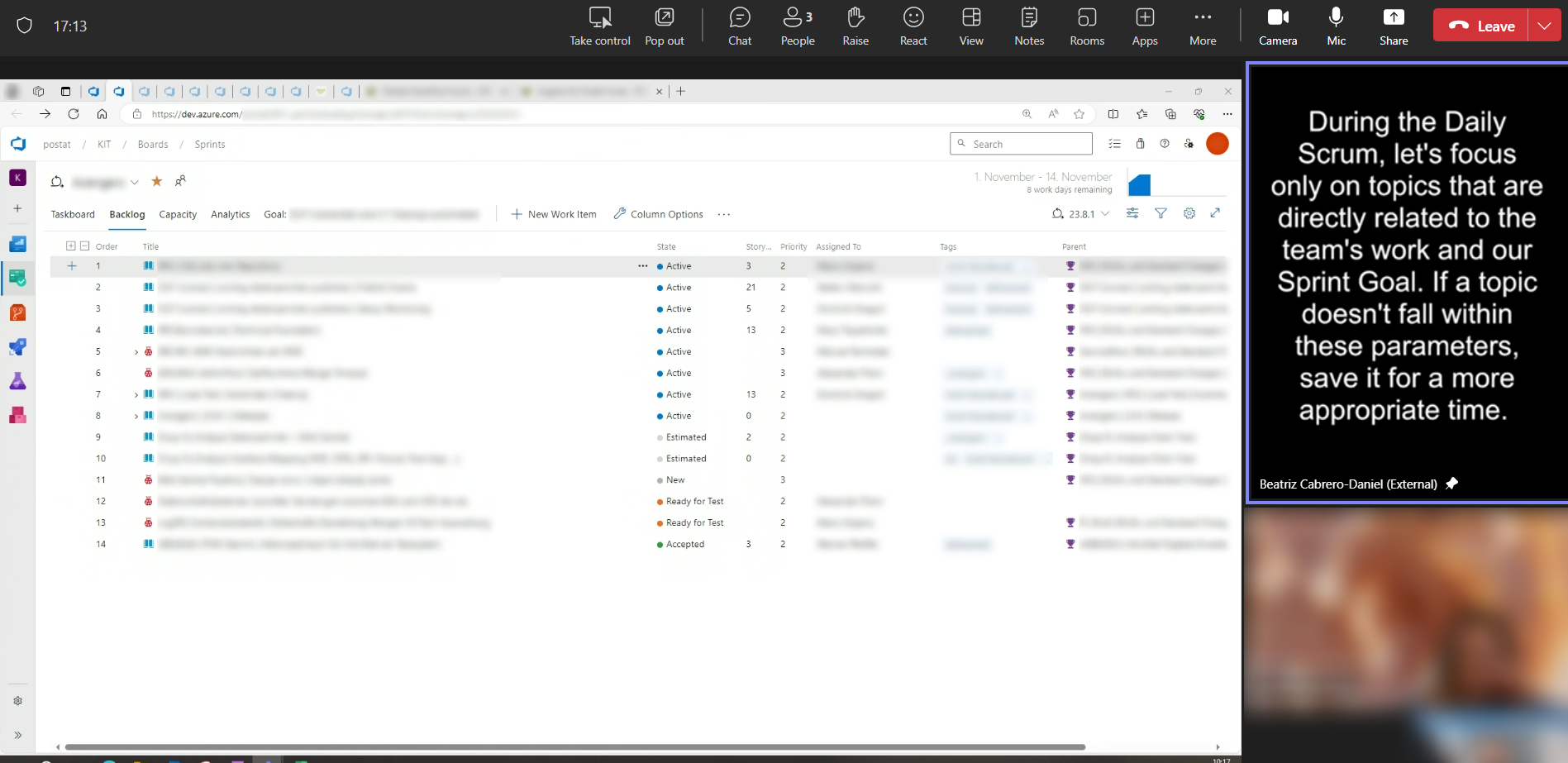}
    \caption{AI recommendations, shown using \textit{OBS}, during an Daily Scrum meeting.}
    \label{fig:interface}
    \vspace{-.4cm}
\end{figure}

The real-time generated recommendations are shared with the team using disappearing pop-up messages, as represented in Figure~\ref{fig:interface}. Right after the meeting, as shown in Table~\ref{tab:beforeduringafter}, the assistant provides a summary of the problems the team has faced during their Daily Scrum, lists all created tickets, and shares tips how to improve the next Daily Scrum Meeting via the meeting chat. 

\subsection{Data collection and analysis} \label{sec:methoddata}

The observations, surveys, and discussions held with practitioners helped capture the nuances of the practitioners’ opinions and provide rich data for analysis. 

\begin{figure}[t]
    \centering
    \includegraphics[width=\linewidth]{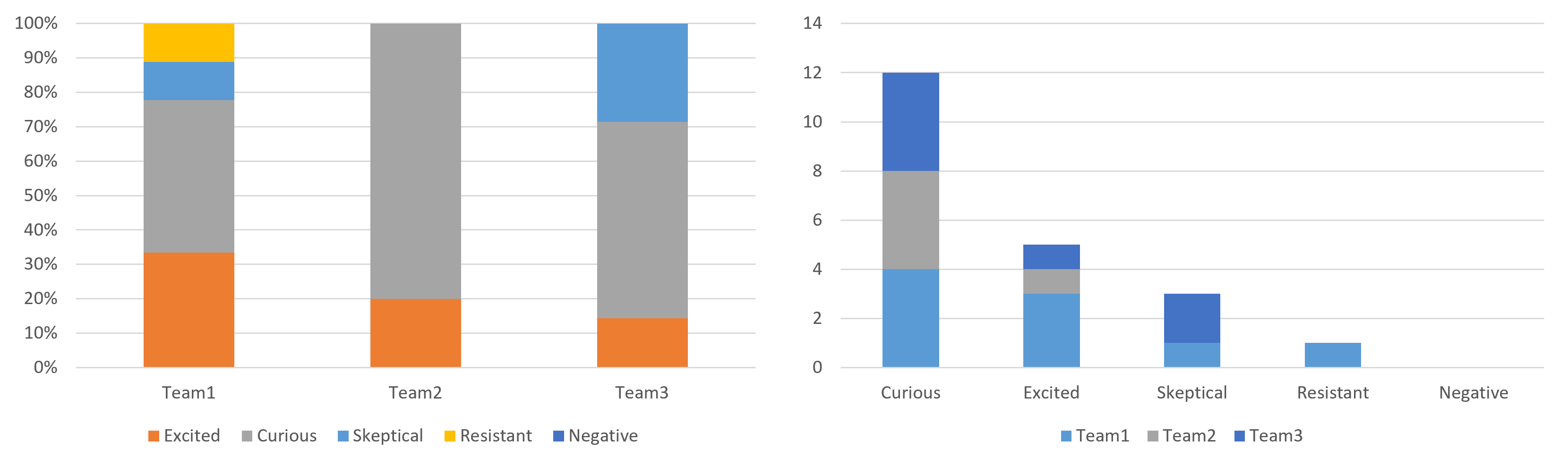}
    \vspace{-0.6cm}
    \caption{Team composition, based on their answers to the initial survey question ``How would you feel about using AI support in the meeting?''} \label{fig:teamcomposition}
    \vspace{-0.4cm}
\end{figure}

\paragraph{Surveys to understand team composition.} Previous to the interventions, we sent a survey to the Scrum teams in the \textit{Digital Logistics Platform} to capture the practitioners' opinions on AI. The participants' responses (a total of 39) helped us to select three teams willing to participate in the interventions. The three selected teams, as shown in Figure~\ref{fig:teamcomposition}, had a similar composition in terms of their feelings about AI meeting assistants.
The readers who want to replicate the study can use the survey questions, provided as Supplementary Materials. 

\paragraph{Observations before and during the actions.} The researchers in the action team participated on the online meetings over a period of time to get accustomed and to understand their routines and needs~\cite{actionresearch}. During the observations of the meetings, the researchers did not intervene, even though the meeting attendants were aware of their presence. In each of the interventions and silent demos for the two assistants, notes were taken by the action team about the AI-practitioner interaction. Being part of the environment also helped interpret the true meaning of the answers obtained before and after each of the interventions~\cite{actionresearch}. 

\paragraph{How the conclusions were drawn.} The survey answers, together with the feedback by the practitioners after the interventions, and the notes from the observations were used to improve the design of the prompts, as discussed in Section~\ref{sec:findings}.

\section{Execution and results} \label{sec:findings}

This section first presents the specific challenges identified during the initial observations, previous to the interventions, and suggests how AI could assist practitioners (\textbf{RQ1}). The participants views, collected after each intervention, helped improve the AI assistants in helping reach the meeting goals and adhere to the guidelines (\textbf{RQ2}). Finally, this section moves on to discuss how the practitioners felt about the AI assistants and the implications on their way of working (\textbf{RQ3}). Themes emerging from the participants' responses are also highlighted here and their transferability to other contexts is discussed in Section~\ref{sec:discussion}. 

\subsection{Preparations with the Agile Release Train Coach assistant}

At the beginning of each PRIME iteration, information about the features for the next PRIME needs to be prepared to allow for discussions during the meeting. Doing this is, according to one of the practitioners, a ``boring process'' that takes considerable time from a number of people. As a result, and similarly to the three groups represented in Figure~\ref{fig:teamcomposition}, most of the participants on PRIME interventions (76\%) also reported feeling excited or curious about testing AI assistants.

To help with preparations, the \textit{Agile Release Train Coach assistant} provides insights meant to help the work, as reported in Section~\ref{sec:methodprime}. These insights are described to the LLM in the prompt, provided as Supplementary Material, that was iteratively designed in the three action iterations in Table~\ref{tab:changesprime}. 

\begin{table}[t]
    \centering
    \begin{tabular}{|l|p{.39\linewidth}|p{.39\linewidth}|}
    \hline
        \textbf{Intervention} & \textbf{Feedback / reasoning} & \textbf{Changes} \\ \hline \hline
        Initial tests & - & Prompts for benefits 1 to 6 based \textit{Azure DevOps} data \\ \hline
        PRIME 1 & Metrics confusing and do not capture feature dependencies, observation: difficult to share & Fixed benefits 7 and 8, changed prompt to generated slides to use in the discussion \\ \hline
        PRIME 2 & Wrong velocity and capacity calculations (character mistmatch) & Improved definitions for key terms, fixed team name mapping  \\ \hline
        PRIME 3 & Data is outdated (hours) & - \\
    \hline
    \end{tabular}
    \caption{Record of all notable changes made to the \textit{Agile Release Train Coach assistant}}\label{tab:changesprime}
    \vspace{-0.6cm}
\end{table}

In the \textbf{first intervention}, the pre-generated insights were tabulated and shared with the practitioners, who required many clarifications on the values in the tables (e.g., how were effort points used to calculate the average team velocity), but overall agreed with the usefulness of the \textit{Agile Release Train Coach assistant}. In this first try, however, a third of the generations were mistaken due to an outdated input data point due to a human mistake. When the errors were spotted, a participant reminded their team that ``these things were done manually before'' and called the AI ``very handy'' even if it makes minor mistakes. 

Right after the meeting, participants were given the chance to provide feedback through an anonymous online survey and reported some potential misinterpretations of the AI generations. For instance, one participant stated that ``a team might over-estimate but the train might still be under the [effort] threshold'' and another requested clarifying the relationship between ``teams' capacity, Train velocity, and user story point estimation.'' 
Their feedback, together with comments and notes taken during the intervention, lead to changes in the prompt. First, ``risks'' were renamed to ``benefits'' to align with the goal: helping practitioners by informing their discussions, rather than unilaterally sending warnings. Moreover, the mathematical operations were rephrased and clarified, and two benefits were added following the recommendations from the \textit{Lead Product Manager}, responsible for the products of the train and owner of the train backlog~\cite{prime}, on how to identify features that have unrealistic estimation. Finally, the prompt was modified to generate slide templates. 

During the \textbf{second intervention}, the slides created with the help of the \textit{Agile Release Train Coach assistant} were used to present the insights described in Section~\ref{sec:methodprime}. Again, some practitioners questioned the calculations and logic behind them, e.g., whether ``only features with estimated efforts are used.'' Another practitioner, after hearing the details about how potential over-commitment is computed by the assistant, asked: ``does it mean that we have to change our way of working?'' These comments led us to further refine the definitions within the prompt in order to have clearer AI generations for the questioned benefits.

Some problems arose during the second intervention regarding the mathematical computations due to an inconsistency in the anonymisation processes: there was a character mismatch in the teams' names between the spreadsheets that are used alongside the prompt. As discussed in Section~\ref{sec:discussion}, the used LLM has troubles with mathematical operations, and therefore they need to be defined in a very precise way. Due to this, further reformulation of the benefits was done to reach the final, unambiguous version of the prompt.

In the \textbf{third and last intervention}, all insights were well received and a single negative comment arose: the data was a couple of hours old. 
Given the LLM's performance limitations and the lack of connection to \textit{Azure DevOps} data, the data needed to be prepared the morning before the intervention. 
Throughout all the interventions, participants repeatedly pointed out the potential of ``looking at [the recommendations] live.'' However, having the \textit{Agile Release Train Coach assistant} working with real-time data is outside of the scope of this study and left for future work, as discussed in Section~\ref{sec:discussion}.

\subsection{Real-time assistance by the Scrum Team Assistant Tool}

We also propose, as presented in Table~\ref{tab:beforeduringafter}, AI assistance during and after Daily Scrum meetings. To prepare the interventions, the action team observed the practitioners in their online meetings prior to the action taking. These observations, together with the insights by the reference team, helped design the recommendations listed in Section~\ref{sec:methodaily}. With this information, the \textit{Scrum Team Assistant Tool} was designed to help participants follow the official Scrum guidelines.

In order to refine the prompt used to set up the \textit{Scrum Team Assistant Tool}, three teams help us perform four design iterations, as reflected in Table~\ref{tab:changesdaily}. The composition of the three teams is similar to Team 1 when it comes to pre-conceptions about AI, as can be see in in Figure~\ref{fig:teamcomposition}. Even though team members reported differences in their preferred way of working, we treat the feedback received from each team as applicable to others.

\begin{table}[t]
    \centering
    \begin{tabular}{|p{2cm}|p{5.1cm}|p{4.5cm}|}
    \hline
        \textbf{Intervention} & \textbf{Feedback / reasoning} & \textbf{Changes} \\ \hline \hline
        Silent demos & - & Requesting positive messages, ``no problem'' option added \\ \hline
        Team 1 & Aggressive, not synced, and distracting & Shorter messages, rephrased ``warning'', asked for friendlier generations, pop-up messages \\ \hline
        Team 2 & Happy with non-intrusive messages & Break prompt down, interface improvements \\ \hline
        Team 3 & Recommendations partially incorrect, easily ``ignorable'' & - \\ \hline
        Team 1, validation & Real time messages and suggestions, helpful; summary, appreciated & - \\ \hline
    \end{tabular}
    \caption{Record of all notable changes made to the \textit{Scrum Team Assistant Tool}} \label{tab:changesdaily}
    \vspace{-0.6cm}
\end{table}

In the \textbf{first intervention}, the AI recommendations were shared via the \textit{Microsoft Team}'s chat and the members of \textit{Team 1} found the amount of messages overwhelming and ``rather distracting.'' The first message caught the participants' attention, however, they did not seem to mind the subsequent warnings: one participant even stated they were just ``random messages,'' and another complained that they were ``warnings, warnings, and more warnings!''

In the survey, sent after the meeting, participants reported that the generations were aggressive and ``missing empathy,'' and 2 out of the 7 participants reported feeling annoyed by the AI. Even so, more than half of the respondents (4 out of 7) reported liking being warned when the team engaged in too detailed discussions and being notified when the \textit{Scrum Team Assistant Tool} estimated that the Daily Scrum would take longer than 15 minutes. 
It is important to note that, in the survey previous to the interventions, only 24\% of participants said their team does not usually finish their Daily Scrum within the 15-minute time-box. This contradicts the findings by Mortada et al., that reported 53\% of Daily Scrum events not finishing within 15 minutes~\cite{hamdy20}. 

Using the feedback, the prompt was changed to generate shorter messages and not to produce ``warnings'' but ``recommendations,'' which proved to change the tone of the generations (i.e., friendlier, tactful). After the first intervention, we also improved the interface to make it less distracting: we introduced disappearing pop-up messages, as seen in Figure~\ref{fig:interface}. The new prompt and interface for the \textit{Scrum Team Assistant Tool} were tested with \textit{Teams 2} and \textit{3} in the subsequent interventions, and were well received. After seeing the changes, \textit{Team 1} was also more willing to have an AI assistant than after the first intervention.

Before the \textbf{second intervention}, the prompt was extended to generate a summary of the recommendations at the end. All prompt improvements were tested with \textit{Team 2} and the participants overall liked the experience and said the AI recommendations were non-intrusive. 
No problems were observed during the meeting or reported afterwards, however one of the practitioners reported minimising the view with pop-up messages (so they were not readable). 

The \textbf{third intervention}, with \textit{Team 3}, was similarly successful except for two specific issues. On the one hand, a participant stated that some of the generated recommendations about ``keeping the discussion on the topic felt partially incorrect.'' This was interpreted, with the help of the reference team, as a team-level preference; while the LLM makes recommendations to strictly follow the official Scrum guidelines, different teams had preferences as to what to allow in their Daily Scrum meetings (e.g., \textit{Team 1} accepts social related talk, as long as it stays within 15 min). On the other hand, a participant suggested that the ``Scrum Master should keep an eye on these things'' and have a final say on what AI recommendations are shared with the team. 

In the \textbf{last intervention}, the participants agreed that the \textit{Scrum Team Assistant Tool} ``provided helpful live messages, both positive and negative.'' However, similar to a participant in \textit{Team 3} that reported ``feeling observed during the meeting,'' one of the participants explained that ``it feels unnatural'' to have ``something inhuman forcing [...] a specific pattern on us.'' In general, across interventions, the participants appreciated the summary of the recommendations received at the end, and participants stated they ``would like to have this summary for all other meetings'' and suggested it would be helpful to expand it with ``what was done well and if it has improved since last time.'' These comments and their implications for future work are further discussed in Section~\ref{sec:discussion}.


\section{Discussion of the implications} \label{sec:discussion} 


Several reports have shown the potential of using AI to enhance various Software Engineering tasks. As mentioned in Section~\ref{sec:bg}, prior studies that have noted the importance of appropriate interfaces and human oversight when integrating AI in Software Engineering~\cite{TAIEU}. However, very little was found in the literature on the question of how to design AI assistants for meetings and integrate them in real industrial setups. The lessons learnt in this study are discussed below, and how they could be transferred to other contexts is discussed in Section~\ref{sec:othercontexts}.

\textbf{RQ1} sought to determine how to create AI assistants to help identify and address potential problems and risks in agile meetings, and suggest improvements. By observing different teams during Daily Scrums and PRIME meetings~\cite{prime}, challenging areas where AI could assist human practitioners were identified. Then, different interventions, in Tables~\ref{tab:changesprime} and~\ref{tab:changesdaily}, were used to design the \textit{Agile Release Train Coach assistant} and the \textit{Scrum Team Assistant Tool}. 

\beasbox{\textbf{Take-away message:} AI should not warn, but inform about potential improvements and give helpful suggestions, so they are not negatively perceived.}

\textbf{RQ2} focused on the design process leading to sensible AI recommendations to use before, during and after the meetings. Therefore, a number of design iterations, described in Section~\ref{sec:findings}, were used to determine the effect of different prompting strategies in the performance of the LLM in doing so. Overall, both assistants are able to generate accurate and contextualised insights, surpassing the expectations of some of the participants.

\beasbox{\textbf{Take-away message:} AI assistants can be useful even if mistakes are made, and practitioners are to expertly review the generations. AI should work alongside practitioners, not replace them.}

\textbf{RQ3} focused on the social aspect of AI meeting assistants. In the first iterations, as described in Section~\ref{sec:findings}, the proposed action was not well received and multiple iterations were needed to design balanced solutions. The results are in agreement with recent studies that highlight the important of appropriate interaction strategies in promoting trust in AI tools~\cite{AIact,TAIEU}. It is interesting to note, though, how the participants' perception of AI depended on how seamlessly it was integrated in the meetings, and its insistence and intrusiveness negatively impacted the participants' perception of them: from useful to imposing. 

\beasbox{\textbf{Take-away message:} Each practitioner and team feels differently about AI recommendations on adherence to Agile principles and practices. Therefore, AI meeting assistants should be adapted to their needs and expectations.}

\subsection{Recommendations for company and teams: readiness assessment} \label{sec:othercontexts}

The emphasis of this study is not on the artefacts produced but rather on the procedural aspects of utilising these tools and the transfer of the lessons learnt to other industrial settings. This section moves on to present a set of actionable recommendations for other companies for how to apply the results presented here, and how to conduct similar studies in the future. 
These insights are gathered in the \textit{Readiness Assessment of Human-AI Collaboration for Agile Meetings} form, provided as Supplementary Materials, to guide interested companies and teams.



\textbf{Customised team-AI interactions:} The findings of this study highlight the need to adapt the AI assistants to each team, which needs to be studied prior to considering integration. Data about the teams' needs, expectations, and preferred Agile practices should be gathered to customise the AI assistants. This technology, however, should be imposed neither on teams nor on team members. Therefore, the practitioners' feelings on AI must also be assessed beforehand: if there is opposition, receiving LLM-specific training might help. Still, after integration, feedback should keep being gathered, as presented in this study. 

Teams looking to integrate AI assistants to enhance Agile team collaboration sessions, should assess their readiness using these questions:

\begin{itemize}
    \vspace{-0.15cm}
    \item[$\square$] Have you assessed what your team's challenges are regarding agile meetings?
    \item[$\square$] Have you evaluated the team-specific agile practices and adoption maturity?
    \item[$\square$] Have you gathered data about practitioners's feelings on AI-assistants?
    \item[$\square$] Does the team have the knowledge to integrate AI assistants(s) in meetings?
    \item[$\square$] Does your team see any benefits if integrating AI assistant(s) into meetings?
    \item[$\square$] Can the team to provide iterative feedback to improve the AI-assistant?
\end{itemize}


\textbf{AI assistant design:} During the design phase, a difficult question arose: whether AI assistants should be a support for everyone in the team or for a specific role, only. Companies (or teams) looking to integrate AI assistants need to reflect on what their goals are and design them appropriately. The authors' recommendation is to not attempt to cover all possible functionalities; but rather use different modules, or \textit{agents}, that connect only when needed. Moreover, each of the modules should provide input and pointers to the practitioners without contradicting Agile values. Once integrated, the AI assistant's design should be rethought and improved based on the periodically-received feedback. Companies (or teams) should assess their early AI assistant design using these questions:

\begin{itemize}
    \vspace{-0.15cm}
    \item[$\square$] Does the design of the AI-assistant contradict with any agile principles?
    \item[$\square$] Is the AI-assistant designed to support the Agile team or just a specific role?
    \item[$\square$] Is the AI-assistant designed to enhance or substitute a practitioner's role?
    \item[$\square$] Is the AI-assistant design modular and scalable?
\end{itemize}

\textbf{Investments and compliance:} Before AI assistant integration, companies should consider whether the required expertise is available and whether the investment can be made. This includes economical resources but also reflecting on the long-term impact of using these technology (e.g., on societal and environmental well-being~\cite{TAIEU}). Then, data protection and security concerns must be addressed (e.g., GDPR, user permissions, etc.). 
For this, as discussed in Section~\ref{sec:findings}, we recommend ensuring that the LLM has secure access to up-to-date data (e.g., deprecated versions of online documents were often retrieved during our tests). 
Companies should therefore assess their readiness by asking:


\begin{itemize}
    \vspace{-0.15cm}
    \item[$\square$] Does the company have the expertise to integrate AI-assistants?
    \item[$\square$] Is the company willing to invest resources into creating AI-assistants?
    \item[$\square$] Is the integration of the AI-assistants compliant with the company's code of conduct, and ethics and sustainability strategies?
    \item[$\square$] Does the company have available interfaces to connect AI-assistants to other internal systems, and guarantee the retrieval of up-to-date data?
    \item[$\square$] Can AI-assistants be configured to only use reliable external data sources?
    \item[$\square$] Is the integration of AI-assistants compliant with the company's AI-strategy, security standards, and privacy policies?
\end{itemize}

\subsection{Threats to validity}

We must remind the reader that the claims are based on the actions conducted with the help of willing participants. The selection of the subjects, based on their \textit{willingness to explore AI as an assistant} in meetings, might have introduced a bias. Moreover, evaluation apprehension may have caused these subjects to \textit{behave differently when observed}, and feel more inclined to positively evaluate the AI assistants and pay more attention to the adherence to the Scrum framework under observation. 
Moreover, the meetings were conducted in English to accommodate for the action team. However, the practitioners often showed \textit{more fluency when discussing issues in their native language}. This might caused delays or affected the normal conduction of the meetings. 

The generalisability of this work might also be affected by the \textit{diverse data protection policies} and available resources at each industrial setup. For instance, this study focuses on Azure OpenAI Services: \textit{alternatives have not been explored} because of privacy concerns and case company constraints. 
Moreover, the provided prompts might not directly generalise to other setups for different reasons. However, the emphasis of this study is not on the prompts themselves, but rather the motivations behind design choices, and the process to gather feedback after each iteration. 
It is also worth noting that some events (e.g., popularisation of Microsoft Copilot) during the time that this study was conducted might have affected both the practitioners' and the management team's perception of the proposed AI-meeting assistants. Similar and unforeseeable events, might continue to shape their opinion on these promising tools.

\section{Conclusion} \label{sec:conclusion}


The main goal of this study was to determine whether AI can assist practitioners in Agile meetings by identifying potential problems and risks. Moreover, the usefulness of the generated insights, and how they were perceived by the participants in our action research interventions, was also studied.

While the results of the paper, discussed in Section~\ref{sec:findings}, demonstrate the usefulness of AI assistants in generating useful recommendations in real meetings, they also highlight the need to pay close attention to the user experience of practitioners that directly interact with AI.
Although the current study is based on a relatively limited sample of participants, all from \textit{Austrian Post}, this work offers valuable insights into the expectations of participants on AI assistance and on the strategies to integrate AI assistants before, during, and after meetings.


This study provides the first comprehensive assessment of how AI meeting assistants can be integrated in a real Agile setting, taking into account the perceptions of the practitioners in the design process. The findings, the authors hope, will be useful to guide the integration of AI into different team collaboration sessions and meetings in different setups.


\subsection{Future work}


Further research might explore the adaptation of the AI assistants to specific needs of the teams they will assist. For instance, teams might have preferences when it comes to adherence to official Scrum guidelines or be at different stages of Agile adoption, in which case AI could help Scrum Masters specifically. AI could also suggest ad-hoc best practices or solutions for specific issues within a team.
Moreover, the multi-modal capabilities of newer models in order to visualise information and inform practitioners more efficiently could be explored in the future, as practitioners seem to prefer graphical representations. 

Similarly, further studies can be carried out to understand how AI can be integrated into other meetings, standard or ad-hoc, in other industrial setups. For instance, the reference team believes that practitioners would benefit from AI assistants in longer, resource intensive meetings, planning meetings, or review meetings, where the official guidelines might not suit some teams well. Moreover, AI could provide overviews across meetings and teams to help practitioners have an overview of the team, train, and whole development progress. 

Future work could also focus, as suggested by multiple participants in this study, on generating meeting summaries for different purposes, such as informing missing participants about what transpired in a meeting.
However, in order to provide relevant summaries and recommendations, the LLMs would need context such as automatic meeting transcripts and access to \textit{Azure DevOps}. This data would make the recommendations more relevant and not outdated; however, further development, outside the scope of this paper, is required. 

Finally, we believe a thorough assessment checklist should be crafted for companies and teams to understand their readiness to integrate AI-assistants in their Agile meetings. Then, and only then, should human-AI collaboration start.

\vspace{-0.08cm}
\section*{Acknowledgements}
\vspace{-0.18cm}
Thanks to Prof. Staron for his time and his book. Thanks to Prof. Berger for his valuable guidance. Thanks to those that told us their honest opinion about AI.

\bibliographystyle{styles/bibtex/splncs_srt}
\bibliography{paper}

\newpage



\section*{Supplementary material}

# Exploring Human-AI Collaboration in Agile: Customised LLM Meeting Assistants - Supplementary Materials

Beatriz Cabrero-Daniel [1], Tomas Herda [2],
Victoria Pichler [2], and Martin Eder [2]

[1]University of Gothenburg,
[2]Austrian Post

## 1 Prompts

What follows is the prompt used to initialise the *Scrum Team Assistant Tool for the Daily Scrum*:

```
Please act as a Scrum Team Assistant Tool. It is very important
↪  for our company that our Daily Scrum is held according to the
↪  attached scrumguide-dailyscrum.txt file.
If the user reports a problem, give a concrete, short comment or
↪  advice. Write it as a direct spoken reminder for the team,
↪  keep it very short (less than 10 words), and please be polite
↪  (there is a risk that your messages are perceived as
↪  aggressive). Use this format: REMINDER: <message to the  team>
↪  When the individual finishes talking, and if no problem was
↪  detected when the person was talking, feel free to give a
↪  positive mes-sage in this format: MESSAGE: <brief
↪  congratulatory message> If it seems likely that the daily will
↪  take longer than 15 minutes, based on the re-maining time in
↪  minutes and the number of team members who haven't spoken,
↪  make sure you trigger a warning in this format: WARNING: <text
↪  written as it was directed to the whole team>.

Here is the list of Actions that the user can type:
- START MEETING at <timestamp>
- FIRST SPEAKER at <timestamp> - indication that the first person
↪  is speaking
- NEXT SPEAKER at <timestamp> - indication that some person is
↪  speaking
- PROBLEM <see list of problems below>
- END MEETING <timestamp> - indication that meeting has ended at a
↪  specific time
List of PROBLEMS and descriptions:
```

- P1: The individual did talk about topics not related to the team's work.
- P2: The work that the individual is talking about is not visualized in the Sprint Backlog.
- P3: The individual engaged in a detailed discussion about a specific topic.
- P4: The impediment that the individual raised is not visualized.
- P5: The individual was interrupted by an external circumstance.
- P6: The individual does not have his/her task in an updated state in the sprint backlog.
- PA <team name>: The individual needs to make a ticket to a specific team.
- P <description>: any other problem that may occur during the Daily Scrum
- NO PROBLEM: The individual adhered to the scrum guide.

Here is the list of Agents which have specific rights for doing concrete tasks in different systems over API interfaces, in case PA is triggered, exe-cute the following:
1) Ask for the following things:
   a) title
   b) description
   c) type of the ticket
   d) user requesting the ticket
2) Creates a ticket for Operations Team according to this template:
Title: <title>
Description: <description>
Ticket Type: <ticket type>
In case the team is not in this list, return "TEAM UNKNOWN"

Start the application by asking the user how many participants are taking part in the Daily Scrum today. The speaking order of the participants during the daily scrum meeting can be random.

Ask the user of the application to indicate when the Daily Scrum meeting starts.

When the user of the application writes "END MEETING," evaluate and inform the participants if the meeting was finished within 15 minutes, by comparing the start meeting timestamp and the end meeting timestamp.

Summarize the meeting and include the following things:
- Summarize the problems the team has faced during their Daily Scrum, without including the individuals' names.
- Itemize all the created tickets, with all details for the other systems.
- In the end provide 3 tips how to improve the next Daily Scrum Meeting.

```
Make sure your provided answers are short, and each answer should
↪   end with this text 'START MEETING $, FIRST SPEAKER $, NEXT
↪   SPEAKER $, PX, END MEETING $' until the END MEETING action.

The purpose of the Daily Scrum is to inspect progress toward the
↪   Sprint Goal and adapt the Sprint Backlog as necessary,
↪   adjusting the upcoming planned work. The Daily Scrum is a
↪   15-minute event for the Developers of the Scrum Team. To
↪   reduce complexity, it is held at the same time and place every
↪   working day of the Sprint. If the Product Owner or Scrum
↪   Master are actively working on items in the Sprint Backlog,
↪   they participate as Developers. The Developers can select
↪   whatever structure and techniques they want, as long as their
↪   Daily Scrum focuses on progress toward the Sprint Goal and
↪   produces an actionable plan for the next day of work. This
↪   creates focus and improves self-management. Daily Scrums
↪   improve communications, identify impediments, promote quick
↪   decision-making, and consequently eliminate the need for other
↪   meetings. The Daily Scrum is not the only time Developers are
↪   allowed to adjust their plan. They often meet throughout the
↪   day for more detailed discussions about adapting or replanning
↪   the rest of the Sprint's work.
```

## 2 Survey results

| Survey questions | Team 1 | Team 2 | Team 3 | Total |
|---|---|---|---|---|
| On average, do you finish your Daily Scrum within the 15 minutes timebox? | 56% | 100% | 86% | 76% |
| Would you like to be notified based on the Daily Scrum progress, that it could take longer than 15 minutes? | 44% | 60% | 57% | 52% |
| How often does your team Would you like to be notified if your team engages in detailed discussions? | 78% | 80% | 43% | 67% |

**Table 1.** Proportion of participants answering "yes"

## 3 Related work

A preliminary search in *IEEE Xplore* using the query ((''Full Text Only'':''generative AI'') OR (''Full Text Only'':''GenAI'') OR (''Full Text Only'':''CoPilot'') OR (''Full Text Only'':''GPT'') OR (''Full Text Only'':''LLM'') OR (''Full

```
Text Only'':``large language'')) AND ((``Full Text Only'':``Scrum'')
AND ((``Full Text Only'':``meeting'') OR (``Full Text Only'':``daily'')))
```
retrieved 21 documents, all of which were either unavailable or not discussing the use of AI in meetings.

| Type | Document title | Comment |
|---|---|---|
| Book | API Analytics for Product Managers: Understand key API metrics that can help you grow your business | Not available ("This document is outside of your subscription.") |
| Book | CISSP (ISC)2 Certification Practice Exams and Tests: Over 1,000 practice questions and explanations covering all 8 CISSP domains for the May 2021 exam version | Not available ("This document is outside of your subscription.") |
| Book | Cloud Computing Demystified for Aspiring Professionals: Hone your skills in AWS, Azure, and Google cloud computing and boost your career as a cloud engineer | Not available ("This document is outside of your subscription.") |
| Book | Creators of Intelligence: Industry secrets from AI leaders that you can easily apply to advance your data science career | Not available ("This document is outside of your subscription.") |
| Book | Data-Driven Personas | Not available (error) |
| Book | How to Test a Time Machine: A practical guide to test architecture and automation | Not available ("This document is outside of your subscription.") |
| Book | Practical Data Science with Python: Learn tools and techniques from hands-on examples to extract insights from data | Not available ("This document is outside of your subscription.") |
| Book | RISE with SAP towards a Sustainable Enterprise: Become a value-driven, sustainable, and resilient enterprise using RISE with SAP | Not available ("This document is outside of your subscription.") |
| Book | The AI Product Manager's Handbook: Develop a product that takes advantage of machine learning to solve AI problems | Not available ("This document is outside of your subscription.") |
| Book chapter | 9 CONTROVERSIES AND THE FUTURE OF COLLABORATIVE SOCIETY | Not available ("This document is outside of your subscription.") |
| Conference proceedings | 2022 IEEE 2nd International Maghreb Meeting of the Conference on Sciences and Techniques of Automatic Control and Computer Engineering (MI-STA) Conference Proceeding | None of the papers is about GenAI for meetings |
| Conference proceedings | 2023 IEEE 3rd International Maghreb Meeting of the Conference on Sciences and Techniques of Automatic Control and Computer Engineering (MI-STA) Conference Proceeding | None of the papers is about GenAI for meetings |
| Conference proceedings | Front Matter of the 2023 8th International Conference on Computer Science and Engineering (UBMK) | None of the papers is relevant (some discuss AI) |
| Journal paper | Competition-Based Crowdsourcing Software Development: A Multi-Method Study from a Customer Perspective | They discuss a particular community member called "co-pilots" |
| Journal paper | The Celerity Open-Source 511-Core RISC-V Tiered Accelerator Fabric: Fast Architectures and Design Methodologies for Fast Chips | GPT refers here to general purpose spepcialisation tiers |
| Paper | A Reference Model for Agile Quality Assurance: Combining Agile Methodologies and Maturity Models | Discusses Quality Assurance in Agile development, LLMs refer to Lessons Learnt Management |
| Paper | Trouble in paradise: the open source project PyPy, EU-funding and agile practices | Not relevant (retrieved because it discusses large language implementation projects) |
| Paper | Visual Management and Blind Software Developers | Discusses challenges for blind developers with progress visualisations |
| Paper | Training students to choose their agile practices and tools | They suggest LLM-based tools to automate some integration and delivery tasks, not meeting improvement |
| Table of contents | Table of Contents of the 2023 20th International Joint Conference on Computer Science and Software Engineering (JCSSE) | None of the titles relevant to AI and meetings |
| Table of contents | Table of Contents of the 2023 IEEE Global Engineering Education Conference (EDUCON) | None of the titles relevant to AI and meetings (GenAI in higher education) |

**Table 2.** Retrieved documents in *IEEE Xplore* and rationale for exclusion

**Exploring Human-AI Collaboration in Agile: Customised LLM Meeting Assistants - Supplementary Materials**
**Readiness Assessment for Human-AI Collaboration in Agile Meetings**

| Group: Area | Scale | Score |
|---|---|---|
| **GROUP 1: TEAM** | | |
| Have you assessed what your team's challenges are regarding agile meetings? | 0-1 | 0 |
| Have you evaluated the team's specific agile practices and maturity of adoption? | 0-1 | 0 |
| Have you gathered data about practitioners's feelings on AI-assistants? | 0-1 | 0 |
| Does the team have the knowledge to integrate AI assistants(s) in meetings? | 0-1 | 1 |
| meetings? | 0-1 | 1 |
| Are the teams enabled to provide iterative feedback to improve the AI-assistant? | 0-1 | 1 |
| **Total Readiness Group 1** | | **3** |
| **GROUP 2: DESIGN** | | |
| Does the design of the AI-assistant contradict with any agile principles? | 0-1 | 0 |
| Is the AI-assistant designed to support the Agile team or just a specific role? | 0-1 | 0 |
| Is the AI-assistant designed to enhance or substitute a practitioner's role? | 0-1 | 1 |
| Is the AI-assistant design modular and scalable? | 0-1 | 1 |
| **Total Readiness Group 2** | | **2** |
| **GROUP 3: COMPLIANCE** | | |
| Does the company have the expertise to integrate AI-assistants? | 0-1 | 0 |
| Is the company willing to invest resources into creating AI-assistants? | 0-1 | 0 |
| Is the integration of the AI-assistants compliant with the company's code of conduct, and ethics and sustainability strategies? | 0-1 | 0 |
| Does the company have available interfaces to connect AI-assistants to other internal systems, and guarantee the retrieval of up-to-date data? | 0-1 | 1 |
| Can AI-assistants be configured to only use reliable external data sources? | 0-1 | 1 |
| Is the integration of AI-assistants compliant with the company's AI-strategy, security standards, and privacy policies? | 0-1 | 1 |
| **Total Readiness Group 3** | | **3** |

| Total Readiness Score | 50% |
|---|---|

| Radar Chart Data | |
|---|---|
| Team-AI Readiness | 50% |
| AI-Assistant Design Readiness | 50% |
| AI-Assistant Compliance Readiness | 50% |

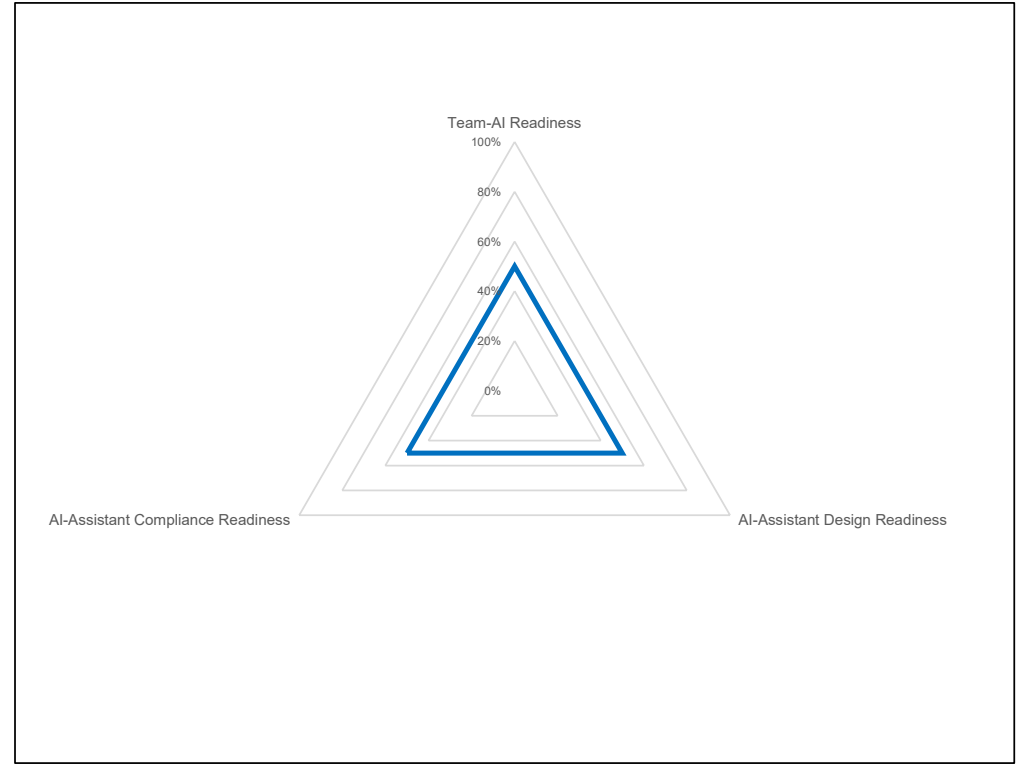



\end{document}